___

# The Dual Imperative: Innovation and Regulation in the AI Era
___

**Innovation and Regulation – a False Choice**

Paulo Carvão
paulo_carvao@harvard.edu

Senior Fellow, Mossavar-Rahmani Center for Business and Government
Harvard Kennedy School
79 JFK Street, Cambridge, MA 02138, USA

**Abstract**

This article addresses the societal costs associated with the lack of regulation in Artificial Intelligence and proposes a framework combining innovation and regulation. Over fifty years of AI research, catalyzed by declining computing costs and the proliferation of data, have propelled AI into the mainstream, promising significant economic benefits. Yet, this rapid adoption underscores risks, from bias amplification and labor disruptions to existential threats posed by autonomous systems. The discourse is polarized between "accelerationists," advocating for unfettered technological advancement, and "doomers," calling for a slowdown to prevent dystopian outcomes. This piece advocates for a middle path that leverages technical innovation and smart regulation to maximize AI's potential benefits while minimizing its risks, offering a pragmatic approach to the responsible progress of AI technology. Technical invention beyond today's most capable foundation models is needed to contain catastrophic risks. Regulation is required to create incentives for this research while addressing current issues.

**Key Words**

Artificial Intelligence; AI; Social Media; Tech Policy; Tech Regulation; AI Safety; AI Governance; Innovation; AI Ethics; Technology

**Biographical Notes**

Paulo Carvão is a Senior Fellow at the Mossavar-Rahmani Center for Business and Government at the Harvard Kennedy School, focusing on AI Policy. A former IBM executive, he advises tech startups and explores technology's impact on democracy and the role of entrepreneurship as a vehicle for social mobility.

## Introduction

Artificial Intelligence has the potential to significantly improve personal and business productivity resulting in enhanced economic output and wealth. On the other hand, given its transformative impact, this technology will affect labor markets and can mirror and amplify biases that currently exist in society and the corpora of data it is trained on. In the long term, it might bring safety risks if misused or developed in a way that allows for and facilitates autonomous replication. It is in this context that the merits of AI regulation are debated today. Should AI be regulated and, if so, how to do it in a way that protects innovation?

According to Ben Buchanan, a professor at Georgetown serving in the White House as special advisor for AI, one can summarize modern artificial intelligence in a single sentence: machine learning systems use computing power to execute algorithms that learn from data. This is the AI triad of algorithms (people), data, and computing power. (Buchanan, 2020) Over fifty years of AI research, catalyzed by declining computing costs and the proliferation of data, have propelled AI into the mainstream, promising significant economic benefits. The release by OpenAI in November of 2022 of a chatbot interface to Large Language Models (LLMs) made this technology available and accessible to the large public. ChatGPT, its product instantiation, achieved 100 million monthly active users in two months making it the fastest-growing consumer application until then. Knowledge-based tasks are being automated and, with that, doors to integrating AI into business processes have been opened.

## Benefits and Risks of AI

The use of Artificial Intelligence goes beyond LLMs and GPTs including other machine and deep learning techniques and product formats. The list of personal productivity enhancements is long. Those include virtual assistants, biometric authentication, personalized news feeds, grammar and spell-checking, smart home devices, navigation apps, fraud detection, and disease prediction based on analysis of personal data. Similarly, there are business value enhancement possibilities. Operations are streamlined, and costs are reduced. Customer experiences are improved through personalized services and support. Companies optimize supply chain management and inventory control and use predictive analytics for better decision-making and market insights. Product and service offerings are upgraded based on customer feedback and data analysis. Education and learning experiences can be personalized and education outcomes can be elevated. Repetitive and mundane tasks are automated allowing humans to focus on more creative work. AI-powered robots can perform dangerous tasks in risky situations that would be unsafe for humans. Some even envision AI helping deliver on the promise of unbiased and accurate decision-making eliminating human biases.

Yet the rapid adoption underscores risks, from bias amplification and labor disruptions to existential threats posed by autonomous systems. This appears to be one of those technologically driven inflection points in history. When this level of change happens, institutions also change, and it is natural to feel a sense of discomfort with the unknown ahead. This unease may not differ from what was experienced around 10,000 BC during the Agricultural Revolution, when the development of agriculture and domestication of animals allowed humans to transition from a hunter-gatherer lifestyle to settled communities, leading to the rise of early civilizations. During the 15th century, the invention of the movable-type printing press enabled the mass production of books, facilitating the spread of knowledge, literacy, and ideas across Europe. In the 18th and 19th centuries, the Industrial Revolution with the development of steam power, mechanization, and factory systems transformed manufacturing processes, leading to urbanization and economic growth. More recently, during the 20th century, computers and the Internet transformed information processing, communication, and access to knowledge on a global scale. At each of these successive waves of technology-driven societal transformation, societies have changed and adapted. They have navigated a narrow path, harnessing the benefits of progress while containing the risks. The journey has not been linear, and there is an ongoing struggle with some of the unintended consequences of this progress, from inequality to climate change, but the pursuit continues.

Something feels different about the change that AI brings. It may be the exponential nature and speed at which it is happening or, using the technology entrepreneur Mustafa Suleyman's characterization "[t]he coming wave is a supercluster, an evolutionary burst like the Cambrian explosion." (Suleyman, 2023, p.57) Or it may be that this one is hitting a bit closer home, threatening the jobs of the elites, and jeopardizing the ability of knowledge workers to earn a living. On a different scale, it prompts the question of what it means to be human if there exists something that can, at some point, approximate or surpass human intelligence. Whatever the reason, it would be a disservice to humankind not to align technology with societal values and use it to advance general welfare. It is at this uncomfortable inflection point that society finds itself, in the middle of a transition and uncertain of the outcome. This creates the conditions for a polarized debate between "accelerationists," advocating for unfettered technological advancement, and "doomers," calling for a slowdown and exhaustive regulation before AI systems are released to prevent dystopian outcomes. The way forward, however, should be a middle path that leverages technical innovation and smart regulation maximizing AI's potential benefits while minimizing its risks, offering a pragmatic approach to the responsible advancement of AI technology. Technical invention beyond the current frontier models is needed to contain catastrophic risks. Regulation is required to create incentives for this research while addressing current issues.

## Types of AI Risks

AI risks typically fall into two main categories. The first encompasses institutional or governance risks. Research in fairness, accountability, transparency, and ethics (FATE) is an increasingly well-established field that views data science and AI as sociotechnical processes, in which the engineering techniques are implemented in a context of social norms and expectations. Scholars in this area document harms including the perpetuation of existing, unjust power relations and the exacerbation of extreme concentrations of power in the hands of a few. (Chan, 2023) They also document problems like privacy violations, evolving vulnerabilities that lead to new security challenges, algorithmic bias, and lack of transparency around how AI models are developed, which can result in accountability and legal issues.

The use of AI by authoritarian governments to control populations poses significant risks to human rights and civil liberties. AI surveillance systems could be used for pervasive monitoring, tracking, and profiling of citizens, enabling unprecedented levels of social control and suppression of dissent. AI decision-making systems could automate discriminatory policies, reinforce biases, and perpetuate injustices. AI propaganda tools could manipulate public opinion and spread disinformation on a massive scale. Overall, AI in the hands of authoritarian regimes could enable new forms of high-tech oppression, eroding privacy, freedom of expression, and democratic principles.

The second category of existential (also known as catastrophic) risks is different. The Center for AI Safety (Hendrycks et al., 2023) grouped these issues into four categories: malicious use, the intentional use of AI to cause harm; AI race, the deployment of unsafe AIs as part of a competitive race; organizational risks, accidents caused human factors or complexity of systems; and rogue AIs, the loss of control to agents more intelligent than humans. These have in common a much larger scale of impact and the fact that they are not likely, at least in their most extreme versions, at the current state of technological development. The pressing question, however, concerns the amount of time available before these risks could materialize, given the accelerating pace of development. Malicious actors could intentionally use AIs to create pathogens for chemical warfare or use AI to lower the cost of propaganda and dissemination of disinformation. The latter is a real and present danger and, the former still lurks on the horizon. Absent adequate incentives and control frameworks, competition between corporations or nation-states can lead developers into a race to release products without the needed safety characteristics. These products can proliferate labor exploitation or lead to large-scale unemployment without the needed societal safety nets. The same technology can be used for development of new autonomous weapons or forms of corporate espionage and cyberwarfare. Lack of focus on safety and unfettered profit-seeking can create an organizational culture prone to accidents. The final group, rogue AIs, captures the popular imagination with super intelligent and agentic AI systems taking over, drifting away from the goals in their original design. While a useful, and theoretically possible, allegory this last scenario is the least probable in the short term.

## The Three Components of AI and Industry Concentration

To understand the current landscape, revisiting the basic building blocks of modern artificial intelligence proves useful: powerful and now affordable computing power is applied to sophisticated algorithms that are trained on massive amounts of data. In their 1943 work *A Logical Calculus of the Ideas Immanent in Nervous Activity*, Warren S. McCulloch a neurophysiologist, and Walter Pitts, a logician, proposed a model of neural activity based on logical operations. This laid the theoretical foundation for modern neural networks by modeling neural activity based on logical operations. (McCulloch & Pitts, 1943) In 1961, at the Massachusetts Institute of Technology, Marvin Minsky in his work *Steps Toward Artificial Intelligence* introduced the AI term. (M. Minsky, 1961) Later, still during this decade, in 1967 the mathematician Alexey Ivakhnenko, in his work *Cybernetics and Forecasting Techniques* introduced the first functional learning algorithm, for which he is often referred to as the father of deep learning. (Ivakhnenko & Lapa, 1967) From there on progress continued with the introduction of backpropagation by scholars at the University of California and Geoffrey Hinton from the Carnegie-Mellon University in the late 1980s (Rumelhart et al., 1986), and a few decades later the seminal *Attention Is All You Need* paper introducing transformers, the base technology for ChatGPT. (Vaswani et al., 2023)

These breakthroughs were followed by periods without much visible innovation, attention, or commercial investments commonly known as the AI winters. This dynamic changed starting in about 2008-2010. By then the cumulative effects of a few decades of Moore's law, doubling the number of circuits in semiconductors every two years while reducing the cost of computing, unleashed a golden era for algorithmic innovation with researchers focused on developing new techniques for AI. People started to question the prevailing idea of diminishing returns on training larger models and started using GPUs' massively parallel processing power to further accelerate the process. The movement of these training runs from private in-house infrastructure to the cloud lowered, even more, the barriers to training models. As an example of the size and scale of contemporary models, GPT-4, the latest language model from OpenAI, released in March 2023, consists of 1.76 trillion parameters.

These models also require massive amounts of data to become useful. The availability of data on the Internet, the exponential increase in data generation by an increasingly connected society, coupled with aggressive data appropriation tactics by tech companies did the trick, and the second ingredient of the AI triad - computing, data, and algorithms - fell into place. The quantity of data used for training is astounding. The most recently released open-source model, Llama 3 from Meta, was trained on up to 15 trillion tokens.

Coming full circle, the considerable size of frontier large models makes them costly to train. Today, estimates suggest that the computing costs alone run into millions of dollars. This represents the largest component of the investment but does factor in the additional expenses for labor and other costs incurred by the development firm. Coupled with these direct costs, there is also material environmental impact given the associated energy demands. Moreover, in their insatiable hunger for data, there is a growing confrontation with the limits of publicly available data scraped from the Internet. This brings attention to the third element of the AI triad: the algorithms and the talent creating them. Arguments are emerging that the limits of current technology with Large Language Models (LLMs) have been reached, and there is a need for faster, leaner, more transparent alternatives with built-in governance. Different innovation streams are working on developing these lighter and more efficient models and, separately, on ways to embed governance and safeguards in the core structure of the models. There may be synergies between the two and simpler models can get us closer to interpretability. For all of these efforts, a limited number of scientists in the world are working at this frontier of science.

Together, these three dynamics are driving industry concentration, with only the best-funded startups surviving and alliances forming between the AI innovation leaders and the largest technology firms and cloud providers. Microsoft's strategic movements in this space are emblematic of these dynamics. From 2019 to 2023, they have invested 13 billion dollars in OpenAI, a significant portion of which in the form of computational resources on Microsoft's Azure cloud service to enable OpenAI's model development and training. During the November of 2023 OpenAI board events - in which their CEO and founder, Sam Altman, was fired and later reinstated - Microsoft further consolidated its influence. In March of 2024, Microsoft hired two of the three co-founders of Inflection AI and most of their 70-person staff. Inflection AI's roots go back to the British startup DeepMind, itself previously acquired by Google. Amazon announced an investment of up to 4 billion dollars in Anthropic as part of a strategic collaboration. These are masterful strategic moves. However, these industry dynamics are leading to a situation where too much power, talent, and resources are concentrated among a handful of players, displaying increasingly monopolistic characteristics.

## Is Governance Keeping Up?

The profoundly transformative potential of AI coupled with concerns about the institutional and catastrophic risks previously discussed has led to the questioning of traditional corporate and state governance models. Startups are experimenting with new models. Governments are scrambling to try and build state capacity and skills to keep up with a new reality. Labor markets are under pressure as even more power shifts from labor to capital.

We can contrast today's corporate governance to the newer models adopted by companies like OpenAI and Anthropic. Traditional corporate governance, from the early days of the 20$^{th}$ century, focuses on shareholder profit maximization. In 1970, economist Milton Friedman argued that a company's sole responsibility is to maximize profits for shareholders while obeying the law, cementing the shareholder model's dominance. (Friedman, 1970) Alternatively, stakeholder governance involves a commitment to value creation, not just for shareholders but also for other stakeholders such as employees, customers, and society at large. Stakeholder governance has seen a resurgence in the past decade as companies are increasingly encouraged to consider their broader societal impact. Benefit corporation laws, which legally require companies to consider stakeholder interests, have been passed in 51 jurisdictions globally since 2010. In 2019, the Business Roundtable, representing 181 CEOs in the United States, issued a statement supporting stakeholder governance over shareholder primacy (Business Round Table, 2019). The 2020 Davos Manifesto by the World Economic Forum advocated for companies to consider their impact on all stakeholders. (Schwab, 2019)

OpenAI's corporate governance structure is uniquely designed to prioritize its mission over profits. OpenAI Global LLC is controlled by a nonprofit, OpenAI, Inc., meaning that investors can't hire or fire board members or control the board. The company's charter emphasizes that OpenAI's mission is to ensure safe artificial general intelligence (AGI) which benefits all humanity, taking precedence over any profit obligations. (Open AI, 2023) This construct was stress tested in November 2023 when the board fired the CEO but reverted this action after pressure from employees and investors. Employee financial considerations over stock option provisions and investors' financial interests prevailed.

Anthropic operates as a public benefit corporation, designed to align long-term societal benefits with its operations, prioritizing responsible AI development over short-term profits. A trust progressively gains authority to, over time, elect a majority of the board members, safeguarding the mission's integrity. Anthropic's Responsible Scaling Policy is central to its governance, establishing detailed safety protocols for AI development. It categorizes risks associated with AI models into different safety levels, stipulating specific actions and checks for each level. This dynamic policy is updated based on evolving AI safety research and best practices and requires halting model scaling if safety standards are breached, maintaining transparency through public reporting, and enforcing uniform safety commitments among all partners. (Anthropic, 2023)

Corporate governance scholar, Roberto Tallarita, highlights several key lessons from the November 2023 corporate turmoil at OpenAI. Despite the attempts at innovative governance, Tallarita notes the persistence of the profit motive, illustrating the complexity of harmonizing it with social objectives. Moreover, independence from profit motives does not necessarily equate to responsible outcomes; active measures are needed to ensure that governance drives towards societal benefits. He also draws parallels between aligning AI's objectives with human values and aligning corporate actions with societal safety. A diverse range of perspectives is crucial to avoid groupthink and ensure thorough risk assessment. Tallarita warns that corporate governance alone is insufficient to mitigate the catastrophic risks posed by AI, calling for robust public oversight and stringent legal frameworks, like those governing other high-risk industries.

(Tallarita, 2023)

## Emerging Regulatory Frameworks

There is a widening gap between the leading edge of technology innovation and the ability to set standards or create regulations. While Artificial Intelligence is not a new science, its popular use is, and generative AI shook the policy world putting into high gear the debate about the ethical and responsible use of the technology. Industry and governments have come up with voluntary standards (The White House, 2023a), the US government issued guidelines (The White House, 2023b), lawmakers enacted regulations (European Parliament, 2023), and libertarians issued manifestos (Andreessen, 2023)  in what looks like a policy free for all. Some say that the guidelines, principles, or voluntary commitments are vague on purpose, to allow room to operate and preempt regulation in an attempt at regulatory capture. Others criticize governments, policymakers, and lawmakers saying that regulating technology is an exercise in futility since it moves too fast. The question as to whether regulating the outcomes is enough and what kind of normative values should be applied by different societies remains open. In this complex system of intertwined questions, much headway has been made during 2023 and we are starting to move away from the ethos of "move fast and break things" to a more nuanced debate about the ethical and responsible use of technology.

Three main emerging approaches are starting to dominate the regulatory scenario. These proposals and laws demonstrate the global race to regulate AI with a goal to balance innovation with managing risks as AI rapidly evolves. The first to be enacted in the form of law is the European Union's Artificial Intelligence Act (EU AI Act). Started in 2021, it was approved by the European Parliament on March 2024 and is now expected to officially become law by later this year. It applies to all 27 EU member states and is likely to create a "Brussels Effect", influencing AI regulation globally. The second approach is the United States federal guidelines, issued via a White House executive order in October of 2023. In the US there is no single comprehensive federal AI law yet, but President Biden has mobilized the regulatory and procurement power of the US government and its agencies. In parallel, Congress is considering AI legislation and some states have passed laws. The Cyberspace Administration of China (CAC), China's top internet regulator, released a draft in April of 2023 of proposed measures for Generative AI to regulate the provision of services within mainland China. Each approach has its advantages and shortcomings and has been designed within the prevailing normative and judicial frameworks of their markets. The European Union emerged as the early regulatory superpower, the United States maintained a tradition of adversarial legalism (Kagan, 2019), and China prioritized state control. These efforts have developed a palette one can now choose from when designing and enacting smart regulation.

In Europe, the "EU AI Act" categorizes AI systems based on their level of risk. The Act prohibits AI systems that pose unacceptable risks and those using manipulative or deceptive techniques, social scoring systems, and certain biometric categorizations, ensuring that these technologies do not undermine fundamental rights or societal values. High-risk AI systems, like those used in critical infrastructure or that influence legal outcomes, face stringent regulatory requirements such as rigorous assessments and compliance processes. Limited-risk AI systems, like chatbots, must ensure transparency to users about their interactions with AI. Most AI applications that pose minimal risk, like video games and spam filters, remain largely unregulated. The AI Act places substantial responsibilities on providers and deployers of high-risk AI systems. Providers, regardless of their location, must meet the Act's requirements if their systems are used within the EU. Deployers, including EU entities using high-risk AI, have specific obligations, however, these are less extensive than those of the providers. The Act also addresses general-purpose AI, mandating that providers, particularly of high-risk or open models, adhere to additional requirements like technical documentation, adversarial testing, tracking and reporting of serious incidents, and cybersecurity measures. (European Parliament, 2023)

In the United States, the Executive Order on the Safe, Secure, and Trustworthy Development and Use of Artificial Intelligence emphasizes a coordinated federal approach to governing AI, acknowledging both its potential and risks. It mandates that AI development adhere to ethical guidelines to ensure safety, security, and trustworthiness, addressing major concerns like discrimination, privacy, and national security risks. Specific measures include guidelines for Federal Government use of AI, enhancing transparency, promoting public and private collaboration, and ensuring robust testing and evaluation mechanisms for AI systems. The order also advocates for global engagement and cooperation to manage AI-related security risks and encourages investment in AI safety research. By integrating these principles, the order aims to harness AI's potential responsibly, prioritizing public welfare and national interests. (The White House, 2023b)

The draft regulation released for comment input by the Cyberspace Administration of China focuses on the provision of generative artificial intelligence services within the country. The measures specify that before offering services using generative AI products, providers must undergo a security assessment approved by the CAC. The regulations also extend to the data used to train or optimize AI products, mandating compliance with China's Cybersecurity Law among other legal frameworks. Such data must not infringe on intellectual property rights and must be accurate, diverse, and obtained legally, particularly when it includes personal information. Furthermore, generative AI service providers are required to ensure users register with their real identities, safeguard user input information, and avoid unlawfully retaining or sharing information that might reveal a user's identity. Violations of these measures could result in penalties as dictated by existing laws like the Cybersecurity Law, Data Security Law, and Personal Information Protection Law, with applicable fines. These measures reflect China's intent to promote the healthy development of AI technologies with an emphasis on strict regulatory compliance. (L. Zhang, 2023)

The regulatory approaches of the European Union, China, and the United States share common objectives with all three prioritizing the safety, security, and ethical development of AI technologies. However, there are notable differences in how these regulations are structured and implemented. The EU's AI Act stands out for its comprehensive risk categorization, proactively setting specific regulations for different levels of AI risk. This contrasts with the more focused approach of China's measures, which are particularly prescriptive about real identity verification and stringent in data handling, indicating a strong emphasis on control and surveillance. Meanwhile, the US Executive Order takes a broader approach, prioritizing national leadership and innovation in AI. It focuses more on setting broad guidelines rather than imposing strict rules, highlighting a preference for flexibility in fostering AI advancement while ensuring safety and ethical standards are met.

As the landscape develops, scholars are starting to provide empirical assessments of how well governments have managed to implement AI governance. A critical view of effectiveness in the US can be seen in *The Bureaucratic Challenge to AI Governance*. (Lawrence et al., 2023) The study illustrates weaknesses in the US government's ability to manage AI governance effectively and highlights challenges stemming from a lack of bureaucratic capacity which hampers effective implementation and strategic planning around AI. *A Roadmap for Governing AI: Technology Governance and Power Sharing Liberalism* studies how AI advancements introduced technologies that draw on vast cultural outputs yet remain poorly understood. These technologies challenge our existing copyright systems, information reliability, power distribution, and the role of technology in democracy. The piece argues for a proactive and expansive approach to AI governance, focusing on human flourishing and democracy, rather than merely reacting to the risks posed by technological advancements. It starts a conversation about reimaging our democratic institutions. (Allen et al., 2024)

The emerging field of regulatory markets offers a promising approach to governance, addressing the limitations of both government-led regulation and industry self-regulation. This model centers on governments defining desired regulatory outcomes—such as safety, fairness, and bias prevention—while private regulators compete to provide solutions that ensure these outcomes are met. By introducing competition, the model encourages innovation in regulatory technologies, allowing the governance of AI systems to keep pace with their rapid development. Traditional regulatory methods often fall short due to the technical complexity of AI, which exceeds the capacity of most government regulators. Regulatory markets respond by shifting the responsibility to private entities, incentivizing them to develop advanced tools and methods for monitoring, auditing, and ensuring AI safety. These market-driven regulators work under government oversight, which ensures accountability while fostering a flexible, adaptive approach to regulation. A critical component of this approach is the design of incentives for auditors, who are encouraged to maintain safety proactively rather than only responding to failures. This shift promotes continuous improvement and innovation in regulatory practices. While regulatory markets offer significant potential, challenges remain, particularly around ensuring sufficient government oversight and avoiding regulatory capture. However, the model represents a dynamic solution for aligning AI development with public safety and ethical standards. (Bova et al., 2024) (Clark & Hadfield, 2019) (Hadfield & Clark, 2023)

Safeguards are necessary to ensure the continued development and deployment of these technologies, either through a new regulatory regime or by enforcing existing laws. The right liability structure must be established to ensure that laws are enforceable in the context of AI technology companies and frontier models. Otherwise, there is a risk of repeating the situation with social media, where large platforms, in the United States, were shielded from liability for user-generated content and operated under strong legal protections. These protections have been a feature of the legal framework in the United States which hosts several of the dominant platforms or represents a very large market opportunity for global players not incorporated in the US like ByteDance's TikTok from China.

With the benefit of hindsight, the harms generated by the previous generation of technologies are now visible. However, it is debatable whether all these problems could have been anticipated and other sociotechnical considerations played a role beyond the US-centric liability shields, explaining why the phenomenon happened across the world. Similarly, there will be challenges in adapting regulations as more is learned about the effects of the technology or if the regulatory design proves incorrect. Outcome-based regulation and resulting liability represent an ex-post regime, acting after potential damage has occurred. It remains to be seen whether this is considered an acceptable risk, relying on the system to self-correct and design fair and safe systems. Furthermore, it is necessary to consider whether there are any categories of risk and use that need to be regulated in advance or even prohibited, as seen in the EU AI Act.

## We Need a Combination of Innovation and Regulation

The dichotomy between technological advancement with no regulation, and those advocating for a slowdown to prevent catastrophic risks is a false choice. Both are needed. It is naïve to assume that corporations, investors, and nation-states will interrupt the progress already made in the field. A balance between technical innovation and smart regulation must be found to maximize AI's potential benefits while minimizing its risks. This balance will come from a combination of scientific and technical inventions that will address the open issues with the current state-of-the-art technology and regulation to address the current institutional risks and create incentives for safety research and development.

Leading techniques in artificial intelligence continue to face substantial technical challenges, notably the alignment problem. This issue revolves around ensuring that AI systems' actions and goals are consistent with human values and objectives. Misalignments can occur when AI systems interpret their objectives in ways that differ from their creators' intentions, often due to the complexity and underspecified nature of their programming. (Ngo et al., 2024) This challenge becomes particularly critical in the context of agentic AI systems. Agentic systems are characterized by their capability to act autonomously. They could accomplish a goal provided by designers, without a concrete specification of how the goal is to be accomplished and, in doing so, they could affect the world without a human in the loop. Multiple agentic systems can initiate action as if they were trained to achieve a particular quantifiable objective and make decisions that are dependent upon one another to achieve a goal or make predictions over a long-time horizon. (Chan et al., 2023) A significant risk with advanced, autonomous AI is its potential to seek power. This encompasses behaviors like acquiring resources, proliferating, and resisting shutdown efforts, which might occur if the AI perceives these actions as necessary to fulfill its misaligned goals. This assumes future AI systems will have advanced capabilities that will outperform humans, will be able to make and execute plans in pursuit of assigned objectives, and will develop strategic awareness that will determine the advantages of assuming and maintaining power over humans and the real-world environment. (Carlsmith, 2022)

Addressing these risks requires robust techniques for containing and guiding AI behavior. The focus must be on creating AI that is not only powerful but also aligned and constrained within safe operational boundaries. Liquid Neural Networks, Objective-Driven AIs, and Generative Flow Networks are examples of innovation trying to address the limitations of LLMs. Liquid Neural Networks (LNNs), developed by a team of researchers led by Prof. Daniela Rus at MIT Computer Science and Artificial Intelligence Laboratory (CSAIL) in 2020, are designed to adapt and learn continuously from streaming data. Unlike traditional large language models, LNNs maintain adaptability post-training, making them suitable for dynamic environments and agentic systems requiring minimal computing resources. LNNs address the black box issue of LLMs by offering greater transparency and safety in decision-making processes. (Hasani et al., 2018) Objective-Driven AIs focus on achieving specific goals, effectively filtering out irrelevant information to enhance decision-making accuracy and efficiency. Their goal is to build machines that will learn as efficiently as humans and animals, and operate at multiple levels of abstraction, enabling them to reason, predict, and plan at multiple time horizons. Joint Embedding Predictive Architecture (JEPA), introduced by Yann LeCun from the New York University and Meta, enhances predictability by separately encoding dependencies between input pairs. Developed around 2020, this architecture addresses challenges in large language models such as transparency and adaptability. In Yann LeCun's vision, JEPA serves as the foundation for Objective-Driven AI by providing a way to learn useful representations from data. The learned representations can then be used by an objective-driven system to reason about the world and plan actions to achieve desired goals. In this paradigm, AI systems are driven by objectives rather than just predicting the future. (LeCun, 2022) Generative Flow Networks (GFlowNets), developed by Yoshua Bengio and colleagues from the Montreal Institute for Learning Algorithms (MILA) at Université de Montréal in 2021, are designed to address large language models' inability to adaptively sample from complex distributions. GFlowNets aim to model the probability of generating data by training to match a predefined reward function. By facilitating better stochastic decision-making and improving sampling efficiency, GFlowNets seek to overcome limitations of LLMs related to data diversity and quality of generated samples. (D. Zhang et al., 2022)

These are three examples of computer science innovation and recent techniques aimed at addressing current technology shortcomings like alignment issues, hallucinations, and lack of ability to control agentic systems. Liquid Neural Networks, Generative Flow Networks, and Objective Driven AIs each present innovative alternatives to traditional Large Language Models, each tailored to specific needs. Liquid Neural Networks offer dynamic adaptability, excelling in handling temporal data but still with significant computational demands. Generative Flow Networks directly learn data distributions, facilitating efficient sampling but struggle with training and scaling issues. Objective Driven AIs excel in specific, targeted problem-solving, though their effectiveness is constrained by the limits of their preset objectives. Common to all, these technologies exhibit trade-offs between specialized capabilities and challenges in generalizability, scalability, and complexity.

There is visibility into what is happening in academic environments, and, to a lesser extent, into activities behind corporate walls. Whether research of this kind will continue to be funded is a matter of policy. Whether it will be adopted in future product releases, is a matter of incentives.

Regulation can play a role in protecting innovation as opposed to stifling it. The realization of any of the catastrophic issues previously discussed would generate significant societal backlash and could prevent the continued enjoyment of the benefits that AI can offer. The three frameworks of AI regulation protect innovation in distinct yet complementary ways. The EU AI Act, by categorizing AI systems based on risk, ensures that lower-risk AI applications face minimal regulatory hurdles, promoting a flourishing environment for technological development while managing more significant risks associated with high-impact AI. The White House Executive Order encourages maintaining US leadership in AI through innovation, fostering a supportive ecosystem for research and development while ensuring safe and ethical AI practices. Meanwhile, China's regulation draft mandates security assessments before deploying AI services, ensuring that innovations are robust and secure, supporting sustainable technological advancement. Each framework balances the need for innovation with necessary precautions to ensure safe and beneficial outcomes.

The EU AI Act, the US Executive Order, and China's draft measures all enhance AI safety incentives by integrating safety into regulatory compliance, national agendas, and foundational service deployment protocols. These frameworks encourage companies to prioritize robust safety measures during AI development and deployment, enabling a responsible AI landscape. In doing so they create financial incentives for companies to invest in safer and more ethical products. This creates a flywheel effect. This type of regulation fosters safety and ethical innovation, which in turn mitigates risks and allows for continued and increased use of the technology. More use attracts more investments which allow continued and safe innovation. Breaking this cycle allowing for catastrophic risk is not in anybody's best interest.

Recent theoretical modeling research on AI governance emphasizes the need for a nuanced balance between innovation and regulation. AI development races often drive competitors to prioritize speed over safety, especially in short-term contexts, increasing societal risks. However, long-term development allows for more strategic risk-taking, potentially fostering greater innovation. One approach proposes voluntary safety commitments, offering flexibility while maintaining accountability through peer sanctions, as an alternative to over-regulation. (Han et al., 2022) Another model highlights the role of discerning users in building trust, suggesting that user engagement and transparency can help condition trust in AI systems. (Alalawi et al., 2024) Additionally, competition-driven dynamics, such as public knowledge of rivals' progress, can exacerbate risky behavior, underscoring the need for managing information in international AI races. (Emery-Xu et al., 2023) Finally, the evolving regulatory frameworks suggest that both incentives and penalties should be tailored to the specific dynamics of the race—whether rapid or extended—to ensure responsible AI development without stifling technological progress. (Han et al., 2020)

**What Have We Learned from Social Media?**

In his 2018 book, *Custodians of the Internet*, Tarleton Gillespie from Microsoft Research, reflects on the previous decade. "The dreams of the open web did not fail, exactly, nor were they empty promises to begin with. Many people put a great deal of effort, time, resources, and dollars to pursue these dreams, and to build infrastructures to support them. But when you build a system that aspires to make possible a certain kind of social activity, even if envisioned in the most positive terms, you also make possible the inverse - activity that adopts the same shape in order to accomplish the opposite end." (Gillespie, 2021, pp. 204-205) In 2022, Joan Donovan, Emily Dreyfus, and Brian Friedberg, wrote *Meme Wars* while at the Harvard Shorestein Center on Media, Politics and Public documenting the informational and cultural wars leading Americans to the point of insurrection on January 6, 2021. Their thoughts rhyme with Gillespie's words when they say that "[p]latforms directly shape the world by sorting, ranking, filing and broadcasting information in different proportions to different groups based upon platform companies' algorithms." Moreover, they posit, "Politics plays an immensely important role in how communication infrastructure is designed, distributed, maintained, and transformed locally, nationally, and globally." (Donovan et al., 2022, pp. 328-329)

In their recent MIT Technology Review article, *Let's not make the same mistakes with AI that we made with social media,* Nathan Sanders and Bruce Schneier, warn against repeating the unregulated growth mistakes of social media with AI technology. Social media, initially lauded for promoting democracy, now faces criticism for spreading misinformation, affecting mental health, and polarizing politics. These issues stem from inherent attributes of social media like targeted advertising, surveillance, virality, lock-in, and monopolization, all of which are present in AI. The unchecked advancement of AI could lead to even more severe societal damages, given its deeper integration into personal and professional lives. The authors advocate for proactive regulation of AI, emphasizing antitrust laws, transparency, oversight, and the promotion of competition. They suggest a government role in developing AI, ensuring it serves public interests over corporate profits. They posit that the need for regulation is urgent, given the significant implications AI holds for communication, learning, lawmaking, and more, stressing that it's crucial to act now to prevent future harm. (Sanders & Schneier, 2024)

Social media issues are complex sociotechnical problems rooted in the economic structures of the Internet. Any solution will require asking the correct technical, social, and public policy questions. Reflecting on a career studying technology and society, the MIT scholar Sherry Turkle in her book *The Empathy Diaries* reminds us that "[k]nowing how to criticize it is becoming more pressing as social media and artificial intelligence insert themselves into every aspect of our lives, because as they do, we are turned into commodities, data that is bought and sold on the marketplace." (Turkle, 2021, p. 347) Many of the issues need to be publicly adjudicated since they involve normative judgments that vary across different communities. Technical solutions may bump into market incentives tied to the business models of the platforms or into liability safe harbors currently embedded in the US judicial system where the largest of these companies are incorporated. Addressing this issue requires nuanced policy interventions that balance economic incentives with the imperative to safeguard the digital information ecosystem.

A combination of regulatory measures, support for technological innovation, transparency, and education can mitigate the present and future risks, without stifling the economic engines of the digital age. This leads back to the central argument of this discussion: that technical invention beyond current frontier models is necessary to contain catastrophic risks, and regulation is needed to create incentives for such research while addressing current issues.

## Conclusion

This article advocates for a path in the evolution of AI that balances technical innovation with necessary regulation. The analysis indicates that advancing AI technology beyond current models is crucial to effectively mitigate catastrophic risks. Simultaneously, implementing thoughtful regulation is essential to incentivize safety-focused research and address existing and emerging concerns.

The argument presented advocates for a dual approach where technological advancement and regulatory measures are intertwined. Effective AI regulation should not stifle innovation but rather guide it in a direction that enhances societal welfare and mitigates risks. The emerging regulatory frameworks from the European Union, the United States, and China demonstrate a collective recognition of the importance of comprehensive governance that accommodates both innovation and safety.

The experiences with other modern technologies, such as social media, underscore the need for a more proactive approach. The lack of adequate social media regulation, extensive legal safe harbors, and the prevailing imbalance between profit and public interest demonstrate the limitations of a laissez-faire approach to technology governance. These lessons highlight that without strategic regulation, the risks associated with technological advancements can undermine their potential benefits. With AI, the benefits are too important for us not to put in place the needed safeguards and incentives for its responsible and safe use.

The choice between halting the development of AI technology and accelerating it without restraint presents a false dichotomy. Ultimately, the future of AI should not be left to market forces or technological determinism alone. It requires a concerted effort from all stakeholders—governments, corporations, civil society, and the scientific community—to craft an environment where technological advancements are aligned with broad societal benefits. This alliance must combine the strengths of cutting-edge technical invention and innovation, robust public policy, and informed social analysis. The right balance between the two will change over time as the technology evolves and potential risks are uncovered. It will also be a normative decision that could vary across different societies based on the ethical, moral, and cultural values and standards that define what is considered rational or optimal behavior in each of these. Emphasizing responsible development, transparency, and inclusivity will ensure that AI serves as a tool for enhancing human capabilities and addressing pressing global challenges rather than exacerbating existing inequalities or creating new forms of risk. As one looks into the future, the collective responsibility is clear.